# Short-time detection of QRS complexes using dual channels based on U-Net and bidirectional long short-term memory

Runnan He[1+], Yang Liu[1+], Kuanquan Wang[1], Na Zhao[1], Yongfeng Yuan[1], Qince Li[1], and Henggui Zhang[1,2,3,4]*

[+]Joint first author

[1]School of Computer Science and Technology, Harbin Institute of Technology, Harbin, China. [2] Biological Physics Group, Department of Physics and Astronomy, University of Manchester, Manchester, UK. [3] Peng Cheng Laboratory, Shenzhen, China. [4] Pilot National Laboratory of Marine Science and Technology, Qingdao, China.

Correspondence:
Henggui Zhang
henggui.zhang@manchester.ac.uk
Qince Li
qinceli@hit.edu.cn

## Abstract

Cardiovascular disease is associated with high rates of morbidity and mortality, and can be reflected by abnormal features of electrocardiogram (ECG). Detecting changes in the QRS complexes in ECG signals is regarded as a straightforward, noninvasive, inexpensive, and preliminary diagnosis approach for evaluating the cardiac health of patients. Therefore, detecting QRS complexes in ECG signals must be accurate over short times. However, the reliability of automatic QRS detection is restricted by all kinds of noise and complex signal morphologies. In this study, we proposed a new algorithm for short-time detection of QRS complexes using dual channels based on U-Net and bidirectional long short-term memory. First, a proposed preprocessor with mean filtering and discrete wavelet transform was initially applied to remove different types of noise. Next the signal was transformed and annotations were relabeled. Finally, a method combining U-Net and bidirectional long short-term memory with dual channels was used for the short-time detection of QRS complexes. The proposed algorithm was trained and tested using 44 ECG records from the MIT-BIH arrhythmia database. It achieved, on average, high values for detection sensitivity (99.56%), positive predictivity (99.72%), and accuracy (99.28%) on the test set, indicating an improvement compared to algorithms reported in the literature.

## 1. Introduction

One of the most widely used noninvasive physiological ways to measure heart activity is the electrocardiogram (ECG). An ECG waveform is characterized by major features such as P waves, QRS complexes, and T waves, each of which reflects the dynamical conduction of excitation waves in the atria and ventricles of the heart. Therefore, abnormal features of the ECG help in diagnosing not only heart disease but also anything else indirectly impacting the activity of the heart [1]. QRS complexes are the most prominent waveforms, as they indicate the ventricular depolarization state. They are usually detected through a computer-aided analysis of ECG signals. As QRS complexes provides rich information about ventricular excitation and conduction, as well as possible ectopic beat condition, for over three decades, attention has been paid to accurately locate the peaks of the R waves, analysis time series of which help in diagnosing diseases such as hypertension in both adults and children [2]. Though there are well established methods to locate R peaks in normal noise-free ECG signals, but it is still a challenge to have a reliable algorithm to identify the different types of QRS complex in noisy and abnormal ECG signals [3]. Hence, an important aspect of current research is to developed an improved method to analyze abnormal and noisy ECG signals.

In recent decades, several algorithms to detect QRS complexes in ECG signals have been presented in the literature. In general, most detection algorithms have been developed in the past few years. A general overview of this subject is presented in [5-7]. Traditionally, the main methods have two essential stages: (1) preprocessing the raw ECG signals; and (2) a decision stage to identify QRS complexes using rules. Normally, the preprocessing involves linear or nonlinear filtering to remove noise and highlight the components of the QRS complexes. The decision stage finds the R peaks [4].

In the preprocessing stage, most of the filtering algorithms enhance the QRS complexes by increasing the signal-to-noise ratio. In decomposing ECG signals, filter banks are employed to form uniform sub-bands for feature extraction [8]. A high-pass filter, MaMeMi, has been used to remove noise from ECG signals [9]. In addition, quadratic and two event-related moving average filters have been used to reinforce the QRS complexes, which aid the detection of R peaks [10, 11]. In all these methods, most of the filtering operations can cancel the noise; however, the amplitude of the R peaks is also reduced, which affects the subsequent decision stage. The major signal processing algorithms suffer from missed or false detections due to the filter bandwidth and the size of the sliding window [12].

The literature about the decision stage is very extensive and many different approaches have been tried. The seminal research by Pan and Tompkins (PT) produced the PT algorithm, which analyzes the slope of the QRS complex [12]. It is widely regarded as one of the gold standard methods. Similar approaches calculate first- and second-order derivatives and then use a sliding window to locate the QRS complexes [13, 14]. These algorithms do not need training datasets, so they are computationally fast; however, they produce many errors due to high-frequency noise [15]. Moreover, they need ideal signals that are free of noise [13].

For the decision stage, other algorithms use signal transforms, such as wavelet transform (WT) and empirical mode decomposition (EMD) [16-30]. WT remove noise through decomposing the original ECG signals into coefficients, which are used to find the QRS complex spectrum. For example, a single-lead detection approach based on WT is proposed in [17]. The multiscale features of a WT have also been adopted to detect QRS complexes [18]. Furthermore, a stationary WT has been applied to extract local minima and maxima from ECG signals [21]. Higher-order statistical features can be extracted from DWT to locate QRS complexes [22]. Although WT have a good time-frequency representation, which is useful for finding QRS complexes, it is difficult to choose the mother wavelet and to determine the level of decomposition. EMD is an alternative way to detect QRS complexes [23]. EMD has been combined with the Shannon energy envelope [24]. EMD does not need to select intrinsic mode functions but is sensitive to noise [30]. Furthermore, Hilbert transforms with sparse derivatives have been used to detect QRS complexes

[15].

Traditional machine learning methods include support vectors machines and neural networks [31-33]. These methods must train a model and moreover, the parameters must be tuned, both of which require many complex calculations. It can be difficult to address effectively the balance between QRS enhancement and noise reduction [34-36].

Compared to traditional machine learning, deep learning is currently a hot research topic. To the best of our knowledge, few researchers have attempted to detect QRS complexes using deep learning. This method has achieved state-of-the-art performance in many fields, such as object detection [37, 38]. Considering the effective application for medical image segmentation, U-Net can be used as an effective method in analyzing ECG signal waveforms. Therefore, to address the shortcomings of those methods mentioned above, such as complex preprocessing and hand-crafting processes, a new automatic feature extraction based on U-Net and bidirectional long short-term memory (BiLSTM) with dual channels is proposed here to detect short-time QRS complexes. After filtering, a dual-channel input with positive and negative ECG signals can better represent the characteristics of the complex information in the signal. With the proposed method, no complicated signal transformation is needed, therefore, reducing the computational cost. Furthermore, the architecture of U-Net can use the available labels more efficiently and it can automatically capture more features from various types of ECG signal, allowing it to locate QRS complexes accurately. After employing bidirectional LSTM to learn context contents, the morphological local features in an ECG signal can be used to enhance the identification of QRS complex regions.

The rest of this paper is organized as follows. In Section 2, the QRS complex detection method is described in detail. In Section 3, the proposed model is trained and evaluated using the MIT-BIH Arrhythmia Database (MITDB). In Section 4, we discuss the limitations of this research and possibilities for future work. Finally, Section 5 concludes our study.

## 2. Method

Fig.1 is a illustration diagram of the proposed QRS complex detection algorithm. It has four steps: (1) filtering and data segmentation, (2) dual-channel processing, (3) detecting QRS complexes with U-Net with bidirectional LSTM, and (4) optimization. In the first step, data segmentation is used to enhance the training dataset. In the second step, the signal is split into two channels (a denoised ECG signal and its inverted mirror ECG signal), which are input into the model in step 3. This is an innovation. The third step combines U-Net and bidirectional LSTM to detect QRS complexes. Finally, the optimization step corrects false and missed detections to increase the detection accuracy. The following sections contain more details of each step.

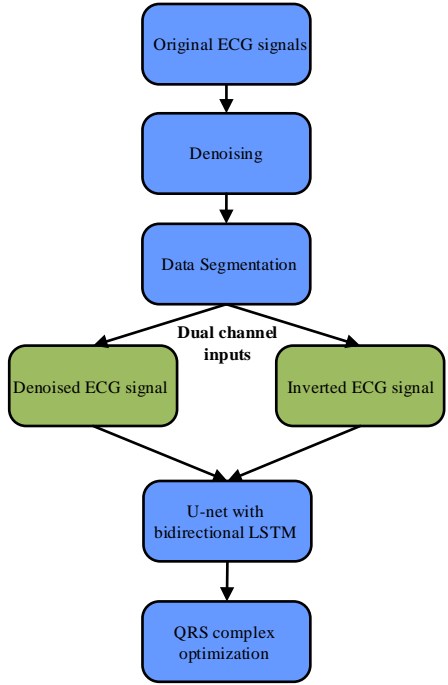

Fig. 1. Illustration diagram of the proposed QRS complex detection algorithm.

## 2.1. Materials

The algorithm was used to detect QRS complexes in 48 records from MITDB [39]. Each record lasts 30 min and was sampled at 360 Hz with 11-bit resolution. A record contains data from modified limb lead II and one of the modified chest leads. The ECG records mainly consist of waveforms and noise, such as artifacts and complex ventricular and conduction abnormalities. For each ECG record, there is an annotation file containing the exact times of when each QRS complex in them occurred. These annotations were added by expert cardiologists. In this paper, the subject-oriented evaluation strategy that was proposed by De Chazal and used by other researchers is applied [40]. According to the Association for the Advancement of Medical Instrumentation (AAMI) [41], four records contain paced beats (records 102, 104, 107, and 217), so these were excluded from the evaluation. The remaining 44 records were divided into a training set and a test set. Each dataset has 22 records (Table I).

TABLE I AAMI-recommended training set and test set used in subjected-oriented evaluation strategy.

| AAMI Classes | Number of records |
| --- | --- |
| Training set | 101, 106, 108, 109, 112, 114, 115, 116, 118, 119, 122, 124, 201, 203, 205, 207, 208, 209, 215, 220, 223, 230 |
| Test set | 100, 103, 105, 111, 113, 117, 121, 123, 200, 202, 210, 212, 213, 214, 219, 221, 222, 228, 231, 232, 233, 234 |

## 2.2. Denoising

ECG signals contain a variety of types of noises including powerline interference and baseline wander. Additionally, large P or T waves can also disturb the detection of QRS complexes. Therefore, as mentioned above, most QRS complex detection algorithms filter the ECG signals in a preprocessing stage [5-11]. In this study, we implement mean filtering to remove baseline wander. Then we apply a five-level DWT to filter out other types of noise using the db4 wavelet. The wavelet coefficients are selected by soft thresholding and then processed by an inverse WT to reconstruct the target signal [42]. Fig. 2 is an example of denoising for ECG record 100. After denoising, the MITDB was divided into 10-s segments, so that the original training set and the test set both have 3960 segments (30 min × 22 records / 10 s).

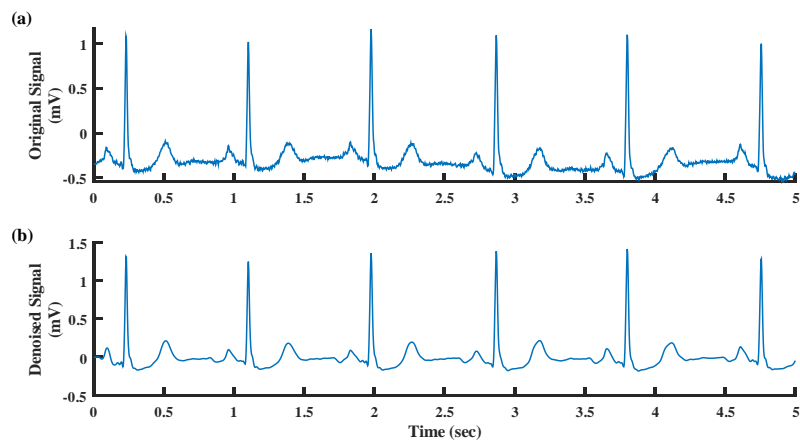

Fig. 2. Removing baseline wander and wavelet denoising. (a) The original ECG signal. (b) The denoised ECG signal.

## 2.3. Dual-channel ECG signals

In the preprocessing step, signal transforms (WT and EMD) were applied to enhance the QRS complex regions [16-30]. However, this increases the amount of computation, which is a problem in short-time QRS complex detection. Thus, we simplified the preprocessing step. Unlike the traditional method, this paper uses deep learning to detect short-time QRS complexes. The inputs to the neural network are signals from two channels. In some conditions of patient's pathology or loss connection of electrode leads, negative R peaks may occur, which reduce the accuracy of detention. To make the algorithm resilience to negative R peaks, we use the inverted ECG signal (i.e., its mirroring image) as a second channel. We input the denoised ECG signal and the inverted ECG signal into the model, as doing so can better represent the context of a ECG signal waveform that is centered on the QRS complex regions and allows the proposed neural network to be trained more effectively. A denoised ECG signal and an inverted ECG signal are shown in Fig. 3.

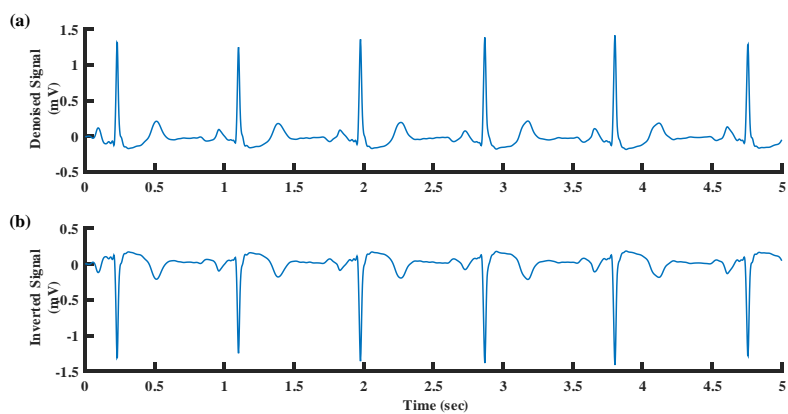

Fig. 3. Example of ECG signal inversion. (a) Denoised ECG signal. (b) Inverted ECG signal.

## 2.4. Decision-making model combining U-Net with bidirectional LSTM

Extracting the locations of features in an ECG signal is an important step in detecting QRS complexes. The morphology of an ECG signal directly affects identification performance. Traditional methods to locate QRS complexes are usually hand-crafted, such as calculating derivates and using moving window integration [12]. In this study, we employ U-Net and bidirectional LSTM to learn the features of QRS complexes automatically from the segmented ECG records.

### 2.4.1 Model input and relabeling the annotations

After preprocessing, we divided the original ECG signals into 7920 segments, each of which lasts 10 s. Each segment was then normalized and centered around zero. Inspired by the application of a deep neural network for image segmentation, the output of our deep neural network is a segmentation map that is the same size as the input. Each value in the map is the probability of an R peak being at that position. This map is a binary matrix where ones denote the position of R peaks and zeros denote anything else, as shown in Fig. 4(b). However, as the R peaks just account for a very small percentage of the sampling points, such an unbalanced classification may lower the accuracy. Thus, we smooth the annotations by converting the unit sample sequence surrounding each R peak into a sequence that gradually changes based on a normal distribution, where the mean ($\mu$) is equal to the R peak index and the standard deviation ($\sigma$) is set to 5 sampling points to fit the QRS width, as shown in Fig. 4(c). Therefore, a sampling point close to an annotated R peak will have a higher probability of being detected as an R peak.

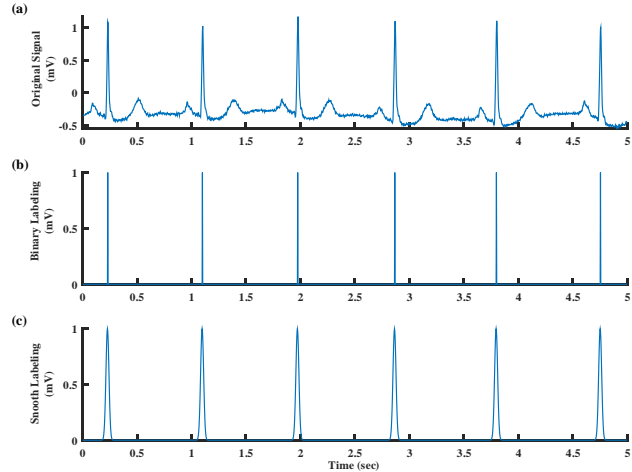

Fig. 4. The relabeling of R peak annotations. (a) Original ECG signal. (b) Binary labels. (c) Smoothed labels.

### 2.4.2 The overall structure

Each ECG signal contains diverse beats. These are regarded as different sequence patterns and they make it difficult to extract QRS complexes. Therefore, a model combining U-Net and bidirectional LSTM was built. In general, U-Net has a contracting path (the left side) and an expansive path (the right side), which are symmetric to capture the context. The contracting path is used to extract high-resolution features and these are merged with the corresponding upsampling output acquired by the expansive path. In this study, unlike a conventional U-Net model [43], before performing the convolution operation on the feature map, we supplement it with a certain edge to make the output feature map consistent with the size of the input. Therefore, the proposed U-Net model does not use cropping operations before concatenating the feature maps. In our U-Net model, the contracting path contains a series of convolutional networks, which consist of the repeated application of two 3 × 1 convolutions (padded convolutions), each followed by a rectified linear unit (ReLU) and a batch normalization layer. There is also a dropout layer between the two convolutional layers to prevent overfitting. After the two convolutional layers, a 3 × 3 max-pooling layer with stride 2 is utilized for each downsampling. At each downsampling step, we double the number of feature maps and halve their size. There are five downsampling stages in total. The expansive path includes the upsampling of the feature maps followed by a 1 × 1 convolutional layer at each step, which halves the number of feature maps. Then, high-resolution features are copied directly from the contracting path and combined with the upsampling features for follow-up convolutions. After that, there are two identical same convolutional layers, as in downsampling. At the final layer, a 1 × 1 convolution is employed to predict the location of QRS complexes.

In addition to the U-Net model, a bidirectional LSTM is used to memorize the context of the input time signal. Because of the gate designs of its unit, the internal state of LSTM can remember the properties of a longer period with up to hundreds of time steps. A bidirectional LSTM has two LSTM layers: the forward LSTM layer and the backward LSTM layer. The latter processes a sequence in the opposite direction. In our design, each LSTM layer has 256 units, which means that the output at each step is a vector with 256 elements. The outputs of the two LSTM layers are summed into a locally focused global feature vector (containing 256 elements), which encapsulates features from the context of the current step in both the forward and backward directions. The outputs of the two LSTM layers are summed stepwise and then fed into the expansive path of U-Net. The overall architecture is shown in Fig. 5.

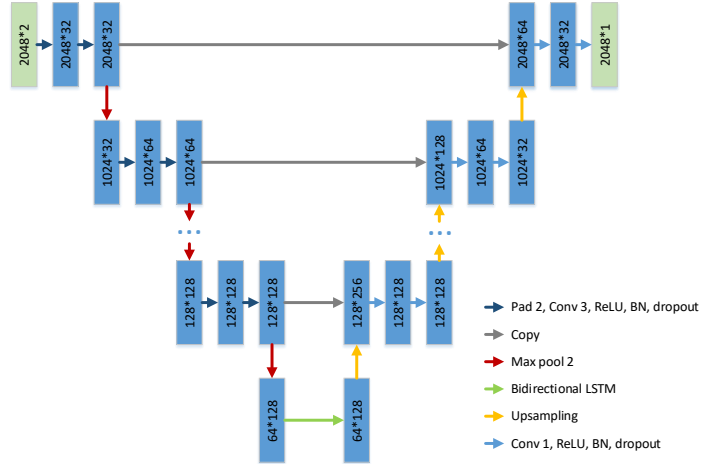

Fig. 5. The network structure for short-time QRS complex detection.

## 2.5. Optimization

The output of the network in this paper is the probability of a sampling point being recognized as the target waveform. An algorithm must determine the position of each QRS complex according to the probability predicted by the model. Therefore, this paper proposes a simple and efficient algorithm whose inputs are the probability sequence output from the neural network and a screening threshold (which is set to 0.1 in this algorithm). The screening threshold is the minimum probability for a sampling point to be predicted as a QRS complex. We can divide a sequence into regions that are below or above the threshold. Then, the maximum in each region above the threshold is identified as a candidate R peak.

According to our experience, if a heartbeat period (RR interval) is at least 1.5 times longer than the average RR interval, then it is possible that some QRS complexes have been missed because their amplitude is too low. Thus, a second threshold that is half the first threshold is applied to these regions to detect any missing QRS complexes. We also set a refractory period of 200ms. If an RR interval is less than 200ms, then there is probably a false detection. In this case, we compare the probabilistic predictions for the two R peaks and eliminate the one with the lower probability.

## 3. Results

The model was built in Python using Keras [44] with TensorFlow as the backend deep learning library [45]. The workstation used for training the models was implemented on a computer with one CPU running at 3.5 GHz, an NVIDIA Quadro k6000 GPU, and 64 Gb of memory.

The neural network was trained and tested on MITDB, which has been widely applied to assess the performance of methods to detect QRS complexes. In this study, the first channel was selected for all 44 records. We calculated three evaluation metrics, as follows. A true positive is when a QRS complex has been detected correctly. For a TP, the detected QRS complex should lie in a region of width 75ms centered on the reference QRS annotation [41]. A false positive is when a QRS complex is reported when no QRS complex is present. A false negative is when a QRS complex is missed. To assess the performance of the proposed algorithm, the sensitivity (*Se*) and the positive predictivity (*+P*) were calculated:

$$Se = \frac{TP}{TP+FN} \times 100\% \quad (1)$$

$$+P = \frac{TP}{TP+FP} \times 100\% \quad (2)$$

where TP is the number of true positives and FN is the number of false negatives. The detection accuracy represents the overall performance of the method:

$$Accuracy = \frac{TP}{TP+FP+FN} \times 100\% \quad (3)$$

where FP is the number of false positives.

Table II shows the results for all 44 records from the MITDB. In total, the proposed algorithm detected 732 FNs and 201 FPs from 101,077 true QRS complexes, such that, on average, Se = 99.28% and +P = 99.80%. For the normal and abnormal ECG signals with different noise levels, the detection accuracy varied from 97.51% to 99.82%. Our low Se of 99.28% may be due to missing small positive R peaks in wide QRS complexes that have large negative S waves.

TABLE II Experimental results for all 44 records from MITDB.

| Record | TP (beats) | FP (beats) | FN (beats) | Se (%) | +P (%) | Accuracy (%) |
|--------|-----------|-----------|-----------|--------|--------|--------------|
| 100 | 2183 | 10 | 12 | 99.45 | 99.54 | 99.00 |
| 101 | 2340 | 5 | 23 | 99.03 | 99.79 | 98.82 |
| 103 | 2235 | 6 | 14 | 99.38 | 99.73 | 99.11 |
| 105 | 2268 | 13 | 17 | 99.26 | 99.43 | 98.69 |
| 106 | 2356 | 5 | 8 | 99.66 | 99.79 | 99.45 |
| 108 | 2291 | 2 | 40 | 98.28 | 99.91 | 98.20 |
| 109 | 2303 | 2 | 12 | 99.48 | 99.91 | 99.40 |
| 111 | 2361 | 6 | 3 | 99.87 | 99.75 | 99.62 |
| 112 | 2252 | 0 | 11 | 99.51 | 100.00 | 99.51 |
| 113 | 2208 | 3 | 9 | 99.59 | 99.86 | 99.46 |
| 114 | 2362 | 1 | 49 | 97.97 | 99.96 | 97.93 |
| 115 | 2345 | 1 | 31 | 98.70 | 99.96 | 98.65 |
| 116 | 2300 | 1 | 10 | 99.57 | 99.96 | 99.52 |
| 117 | 2199 | 7 | 7 | 99.68 | 99.68 | 99.37 |
| 118 | 2278 | 3 | 6 | 99.74 | 99.87 | 99.61 |
| 119 | 2269 | 0 | 7 | 99.69 | 100.00 | 99.69 |
| 121 | 2294 | 2 | 3 | 99.87 | 99.91 | 99.78 |
| 122 | 2287 | 1 | 8 | 99.65 | 99.96 | 99.61 |
| 123 | 2235 | 7 | 12 | 99.47 | 99.69 | 99.16 |
| 124 | 2338 | 4 | 44 | 98.15 | 99.83 | 97.99 |
| 200 | 2244 | 4 | 16 | 99.29 | 99.82 | 99.12 |
| 201 | 2267 | 4 | 64 | 97.25 | 99.82 | 97.09 |
| 202 | 2186 | 5 | 14 | 99.36 | 99.77 | 99.14 |
| 203 | 2279 | 0 | 41 | 98.23 | 100.00 | 98.23 |
| 205 | 2286 | 4 | 42 | 98.20 | 99.83 | 98.03 |
| 207 | 2326 | 3 | 9 | 99.61 | 99.87 | 99.49 |
| 208 | 2431 | 3 | 10 | 99.59 | 99.88 | 99.47 |
| 209 | 2301 | 2 | 6 | 99.74 | 99.13 | 99.65 |
| 210 | 2260 | 5 | 10 | 99.56 | 99.78 | 99.34 |
| 212 | 2224 | 3 | 13 | 99.42 | 99.87 | 99.29 |
| 213 | 2284 | 8 | 22 | 99.05 | 99.65 | 98.70 |
| 214 | 2295 | 12 | 19 | 99.18 | 99.48 | 98.67 |
| 215 | 2337 | 8 | 26 | 98.90 | 99.66 | 98.56 |
| 219 | 2237 | 6 | 5 | 99.78 | 99.73 | 99.51 |
| 220 | 2340 | 1 | 11 | 99.53 | 99.96 | 99.49 |
| 221 | 2283 | 9 | 4 | 99.83 | 99.61 | 99.43 |
| 222 | 2278 | 9 | 13 | 99.43 | 99.61 | 99.04 |
| 223 | 2271 | 7 | 51 | 97.80 | 99.69 | 99.35 |
| 228 | 2198 | 3 | 8 | 99.64 | 99.86 | 97.51 |
| 230 | 2316 | 3 | 3 | 99.87 | 99.87 | 99.74 |
| 231 | 2241 | 1 | 3 | 99.87 | 99.96 | 99.82 |
| 232 | 2194 | 8 | 3 | 99.86 | 99.64 | 99.52 |
| 233 | 2294 | 11 | 5 | 99.78 | 99.52 | 99.31 |
| 234 | 2269 | 3 | 8 | 99.65 | 99.87 | 99.52 |
| Overall | 100,345 | 201 | 732 | 99.28 | 99.80 | 99.28 |

### 3.1. Generalization

The proposed algorithm can generalize well from the training set without additional network regularization. As shown in Table III, generally, the results for the test set are better than those for the training set, which demonstrates that the model is not overfitting. The good generalization from combining U-Net and LSTM may be due to several factors. First, the datasets used for training and testing the model were complex and diverse due to the segmentation. Second, the reduction in the number of learnable parameters mitigated the risk that the model would overfit. This was achieved by using a relatively small kernel in the U-Net model.

TABLE III Experimental results for the training and test set.

| Dataset | TP | FP | FN | Se (%) | +P (%) | Accuracy (%) |
|---|---|---|---|---|---|---|
| Training set | 50,875 | 60 | 512 | 99.00 | 99.79 | 98.89 |
| Test set | 49,470 | 141 | 220 | 99.56 | 99.72 | 99.28 |

### 3.2. Dual channels

The performance of the method with one channel was compared to the performance with two channels (Table IV). Overall, using the dual-channel input is better than using a single channel, which proves its effectiveness. In addition, the advantage of the dual-channel input is illustrated by several representative examples (Figs. 6, 7, and 8). The blue stars in the figures are the annotation of the R peaks, whereas the blue and red dots mark the R peaks detected with dual- and single-channel inputs, respectively.

TABLE IV Experimental results for single- and dual-channel inputs.

| Number of channels | TP | FP | FN | Se (%) | +P (%) | Accuracy (%) |
|---|---|---|---|---|---|---|
| One | 49,446 | 144 | 244 | 99.51 | 99.71 | 99.22 |
| Two | 49,470 | 141 | 220 | 99.56 | 99.72 | 99.28 |

As shown in Fig. 6, with a single channel as input, a peak was missed for the second QRS complex. The reason may be because the local amplitude of the ECG signal was very high for the second QRS complex. With two channels, the inverted signal was used as the auxiliary input. This increases the diversity of the input so that model learns the variation in the QRS complexes better.

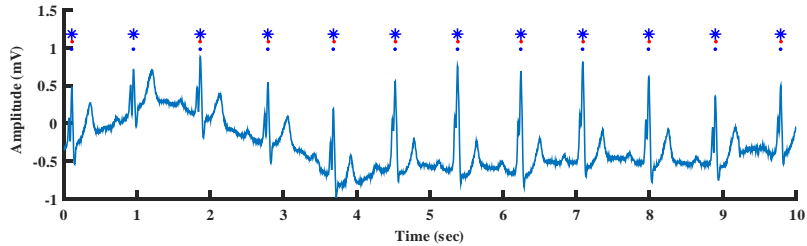

Fig. 6. Comparison of detections with single- and dual-channel inputs (single missed R peak). The blue stars are the annotation of the R peaks, whereas the blue and red dots mark the R peaks detected with dual- and single-channel inputs, respectively.

Fig. 7 is an example where using a single channel as an input, there is one false detection in the middle of the signal (at around 4 s). Although the difference of the number of false detections with single-channel and dual-channel inputs is only three, as shown in Table III, the number is negligible compared to the total number of QRS complexes. Note that the false detection is obviously not an R peak, but the next minimum after a QRS complex. According to an empirical analysis, this error should not occur. In model learning, this sampling point was regarded as an inverted QRS complex, because many ECG signals are anomalous due to human error or various arrhythmias, which can result in inverted QRS complexes. Detecting inverted QRS complexes is a common problem. Because the dual-channel input has an inverted signal, model learning can better extract the features of QRS complexes.

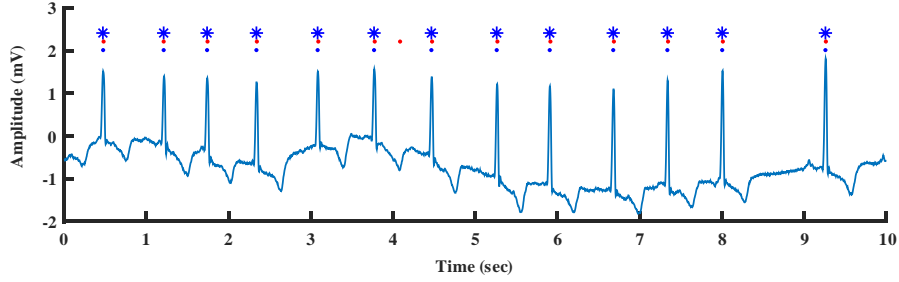

Fig. 7. Comparison of detections with single- and dual-channel inputs (single false R peak). The blue stars are the annotation of the R peaks, whereas the blue and red dots mark the R peaks detected with dual- and single-channel inputs, respectively.

The last example, shown in Fig. 8, has three missed detections. They all occur when the local amplitude of the ECG signal is lower. It is the opposite of the example shown in Fig. 6. When the amplitude of the ECG signal is locally lower or higher, the model learning does not work well. Therefore, using dual channels as input can avoid missing R peaks due to local changes in the amplitude of the ECG signal.

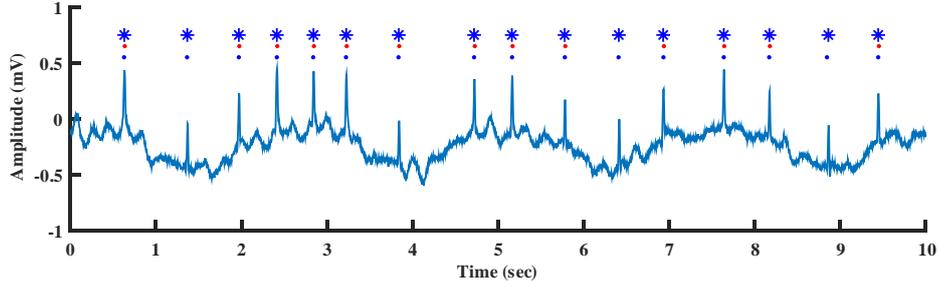

Fig. 8. Comparison of detections with single- and dual-channel inputs (multiple missed R peaks). The blue stars are the annotation of the R peaks, whereas the blue and red dots mark the R peaks detected with dual- and single-channel inputs, respectively.

### 3.3. Smooth labeling

To evaluate the effectiveness of smooth labeling, we compared the performance of models using either binary or smooth labeling. The results are shown in Table V. The number of false positives and the number of false negatives (141 and 220, respectively) for the model with smooth labeling are both significantly less than for the model with binary labeling (557 and 687, respectively). Thus, the model with smooth labeling has a higher Se and a higher +P. One possible reason is that smooth labeling alleviates the imbalance between R peaks and the other sampling points. Besides, the smooth labeling is not very sensitive to the few errors in the annotations, so these errors do not induce a large bias in the model.

TABLE V Experimental results for different labeling methods.

| Labeling method | TP | FP | FN | Se (%) | +P (%) | Accuracy (%) |
|---|---|---|---|---|---|---|
| Binary labeling | 49,003 | 557 | 687 | 98.61 | 98.87 | 97.52 |
| Smooth labeling | 49,470 | 141 | 220 | 99.56 | 99.72 | 99.28 |

## 3.4. Using the PT algorithm for short-time detection

The PT algorithm does not perform well with a short-time ECG signal. The results for ECG signals of different lengths are shown in Table VI. This algorithm has two learning phases followed by a final detection phase. The first learning phase needs about 2s of signal to initialize detection thresholds based on signal and noise peaks. The second phase requires two heartbeats to initialize the average RR interval and the RR interval limit. Then, it starts to identify QRS complexes using these thresholds and other parameters. Because of the time needed for initialization, it works poorly with ECG signals only 5s long.

Moreover, the algorithm applies two thresholds to reduce the number of missed beats. These thresholds are based on the nearest detected signal and noise peaks. Therefore, a short-time ECG signal of <1min is not long enough for the algorithm to adjust the thresholds.

In addition, if the PT algorithm cannot detect any QRS complexes in an interval of length corresponding to 166% of the current average RR interval, the maximum detected peak in that interval is regarded as a possible QRS complex. Depending on the average RR interval, two different calculation methods are employed. Both use the most recent eight RR intervals. Therefore, previously detected QRS complexes must be available, so this method is not suitable for detecting short-time QRS complexes.

The proposed model can automatically extract relevant features and identify QRS complexes without prior knowledge.

TABLE VI Experimental results with PT algorithm for different ECG signal lengths.

| Signal length | TP | FP | FN | Se (%) | +P (%) | Accuracy (%) |
|---|---|---|---|---|---|---|
| 5 s | 99257 | 1075 | 2097 | 97.93 | 98.93 | 96.90 |
| 10 s | 99317 | 907 | 1746 | 98.27 | 99.10 | 97.40 |
| 20 s | 99584 | 848 | 1479 | 98.54 | 99.16 | 97.72 |
| 30 s | 99687 | 795 | 1376 | 98.64 | 99.21 | 97.87 |
| 5 min | 99917 | 772 | 1146 | 98.87 | 99.23 | 98.12 |

## 3.5. Comparison with other algorithms

Table VII compares the performance of the proposed algorithm with several other methods on the same test set. The table shows that our proposed method achieves the best performance with Se = 99.56%, +P = 99.72%, and accuracy = 99.28%. We can conclude that the simple and effective preprocessing and decision steps can improve the accuracy of detecting QRS complexes for various normal and abnormal conditions.

TABLE VII Experimental results for various algorithms on the test set.

| Algorithm | TP | FP | FN | Se (%) | +P (%) | Accuracy (%) |
|---|---|---|---|---|---|---|
| Single channel | 49,446 | 144 | 244 | 99.51 | 99.71 | 99.22 |
| Dual channels | 49,470 | 141 | 220 | **99.56** | **99.72** | **99.28** |
| Dual channels without bidirectional LSTM | 49,459 | 162 | 231 | 99.54 | 99.67 | 99.21 |
| PT algorithm [12] | 48,992 | 233 | 698 | 98.60 | 99.53 | 98.14 |
| Fast QRS algorithm [7] | 48,235 | 681 | 1455 | 97.07 | 98.61 | 95.76 |

## 3.6. Combining U-Net with bidirectional LSTM

The combination of U-Net and LSTM is one of the major innovations of this paper. U-Net is a classic type of network. It has been successfully used for image segmentation. By converting the QRS annotations into a sequence with the same length as the ECG signal, we transformed the QRS detection problem into a sequence segmentation problem. Because bidirectional LSTM is good at sequence processing, we extended U-Net by combining it with bidirectional LSTM. To evaluate the advantages or disadvantages of this combination, we ran a control experiment in which only U-Net was used in learning how to detect QRS complexes with the same input and annotations. The results are shown in Table VII. The combination of U-Net and bidirectional LSTM performs better than U-Net alone in terms of Se, +P, and accuracy. One possible reason is that bidirectional LSTM gives the neural network more of a global view when making local predictions. Whereas a QRS complex can usually be recognized in a short segment (maybe 1 or 2s surrounding the QRS complex position), a wider context may be helpful when the shape of the QRS complex is somewhat unusual or the signal is blurred by noise.

## 4. Discussion

QRS complexes are the most significant features in ECG signals. Locating them contributes to finding other components, such as P and T waves [26]. Furthermore, they can provide useful information for computing heart rates, classifying heartbeats, etc. [46,47]. Detecting QRS complexes is important because they are used to diagnose some abnormalities. The algorithm for detecting QRS complexes in short-time ECG signals proposed in this paper is accurate for a variety of ECG signals.

The following points are some of its shortcomings and possibilities for future work:

1) This algorithm locates only QRS complexes. However, it could be extended to identify other ECG signal waveforms, such as P waves, T waves, and ST segments, which are used to diagnose other types of heart disease (such as myocardial ischemia).

2) The input to the algorithm proposed in this paper consists of denoised ECG signals and inverted ECG signals. Although it works very well, it still misses some R peaks and makes false detections. Subsequent research could consider using first-order or second-order differential ECG signals as input to the neural network. Having more comprehensive input information may improve the automatic extraction of network model features.

3) In this work, we used bidirectional LSTM at the bottom of the neural network, where the feature map has the shortest length. Using it in other parts of the network (e.g., before the layer for prediction) may improve the accuracy of detection. However, since bidirectional LSTM has poor parallelism, the processing time will be longer for a longer input sequence. Therefore, we made a trade-off between accuracy and processing time. In future work, we will explore other methods that may complement U-Net and bidirectional LSTM (or other types of recurrent neural network) to increase effectiveness or efficiency.

4) The proposed model works only with ECG signals that are 10s long. The bidirectional LSTM network may perform differently for an input with a different length. Subsequent work could attempt to detect QRS complexes in ECG signals of different lengths without using a bidirectional LSTM network, because if only U-Net is used to detect QRS complexes, then the data length is not important. Therefore, we need to consider how to improve the adaptability of the model to data length changes while ensuring the accuracy of detection.

5) Using more data for training and testing would make the model more robust.

## 5. Conclusion

In this paper, an efficient deep learning algorithm has been proposed for the automated detection of QRS complexes. We implement mean filtering and a five-level DWT to remove different types of noise. We applied a simple ECG signal channel transformation to enhance the detection of QRS complexes. Since the output of the model is the probability of each sampling point being recognized as a QRS complex, the target output of the network is generated based on the labeled position of the target waveform. This process is called annotation relabeling. Finally, the decision stage is based on U-Net and bidirectional LSTM, which is an effective way to locate QRS complexes. MITDB was used for training and testing the performance of the proposed method, which has a high averaged detection sensitivity, positive predictivity, and accuracy of 99.56%, 99.72%, and 99.28% on the test set, respectively. The proposed method is more accurate than existing methods for detecting QRS complexes in ECG signals containing various normal and abnormal QRS complex waveforms and different types of noise.

## Conflict of Interest Statement

The authors declare that the research was conducted in the absence of any commercial or financial relationships that could be construed as a potential conflict of interest.

## Authors' contributions

RH and YL wrote the first draft of the manuscript, designed and implemented the algorithm. HZ and QL contributed to wrote the manuscript and helped with the algorithm. KW and YY commented on and approved the manuscript. NZ participated in the research on the detection of QRS complexes. All authors read and approved the final manuscript.

## Acknowledgments

The work is supported by the National Natural Science Foundation of China (NSFC) under grants 61572152, 61571165, and 61601143 (to HZ, KW, and QL), the Science Technology and Innovation Commission of Shenzhen Municipality under grants JSGG20160229125049615 and JCYJ20151029173639477 (to HZ), and the China Postdoctoral Science Foundation under grant

2015M581448 (to QL).